\def   \ni {\noindent}

\def   \ssk {\vskip  5truept}

\def   \bsk {\vskip 15truept}

\def   \newline {\hfil\break} 

\def\phn{\phantom{0}}

\def\lesssim{\mathrel{\hbox{\rlap{\hbox{\lower4pt\hbox{$\sim$}}}\hbox{$<$}}}}
\def\gtrsim{\mathrel{\hbox{\rlap{\hbox{\lower4pt\hbox{$\sim$}}}\hbox{$>$}}}}

\def\arcdeg{\hbox{$^\circ$}}

\documentstyle[psfig]{article}

\begin{document}

\font\abstract=cmr8
\font\keywords=cmr8
\font\caption=cmr8
\font\references=cmr8
\font\text=cmr10
\font\affiliation=cmssi10
\font\author=cmss10
\font\mc=cmss8
\font\title=cmssbx10 scaled\magstep2
\font\alcit=cmti7 scaled\magstephalf
\font\alcin=cmr6 
\font\ita=cmti8
\font\mma=cmr8
\def\ref{\par\noindent\hangindent 15pt}
\null
%\vskip 3.0truecm
%\baselineskip = 12pt

%%%%%%%%%%%%%%%%%%%%%%%%%%%%%%%%%%%%%%%%%%%%%%%%%%%%%%%%%%%%%%%%%%%%%%%%%%%%%%%
% ------ beginning of font "title" ------

\title{\ni PROBING THE SMALL-SCALE ANGULAR\\ DISTRIBUTION OF
GAMMA-RAY BURSTS WITH\\ COMBINED BATSE/ULYSSES 4B LOCATIONS}

% beginning of font "author and affiliation"
\bsk \bsk
\author{
   \ni R.~M.~Kippen $^{1,2}$, K.~Hurley $^{3}$, and G.~N.~Pendleton $^{1,2}$
   \bsk
   \affiliation{1) University of Alabama in Huntsville, Huntsville, 
   AL 35899, USA

   \affiliation{2) NASA/Marshall Space Flight Center, Huntsville, AL 35812, USA

   \affiliation{3) Space Sciences Laboratory, University of California,
      Berkeley, CA 94720, USA

} 
\bsk
\baselineskip = 12pt

%%%%%%%%%%%%%%%%%%%%%%%%%%%%%%%%%%%%%%%%%%%%%%%%%%%%%%%%%%%%%%%%%%%%%%%%%%%%%%%
% beginning of font "abstract and keywords"
\abstract{\ni ABSTRACT

\ni We investigate the angular distribution of gamma-ray bursts using the
largest catalog of well-localized events that is currently
available---combined BATSE/Ulysses burst locations.  We present the
preliminary spatial analysis of 415 BATSE/Ulysses bursts included in
the BATSE 4B catalog.  We find that the locations are consistent with
large- and small-scale isotropy, with no significant clustering or
repetition.  We also search for cross-correlation between the
BATSE/Ulysses bursts and known extragalactic objects---such as Abell
galaxy clusters, radio-quiet QSO, AGN and supernovae---and place
limits on the fraction of bursts that could originate from these
objects.  }
\bsk
\baselineskip = 12pt
\bsk
\baselineskip = 12pt

%%%%%%%%%%%%%%%%%%%%%%%%%%%%%%%%%%%%%%%%%%%%%%%%%%%%%%%%%%%%%%%%%%%%%%%%%%%%%%%
%-----------------------------------------------------------------------------%

\text

\ni 1. INTRODUCTION
\ssk

\ni Gamma-ray burst (GRB) astronomy has thus far been explored in two
vastly different realms of angular scale.  The large-scale
distribution of GRB sources has been probed in detail by the Burst and
Transient Source Experiment (BATSE), which finds a highly isotropic
scattering of more than 2000 sources.  Conversely, few bursts have
been localized with high-precision.  While these few events have
allowed the recent breakthrough discoveries of low-energy afterglows
and host galaxy redshift measurements, they have not provided much
insight into the small-scale distribution of the GRB source
population.  We summarize results from our on-going program (see
Kippen et al.\ 1998a) investigating the properties of this distribution
using the largest catalog of well-localized events currently
available---combined BATSE/Ulysses (B/U) 4B burst locations.  We also
use this catalog to investigate possible correlations between GRBs and
known extragalactic objects.

%-----------------------------------------------------------------------------%
\bsk
\ni 2. COMBINED BATSE/ULYSSES GRB LOCATIONS
\ssk
\ni The GRB instrument aboard Ulysses is only sensitive enough to
detect the $\sim$25\% most highly fluent BATSE bursts.  However,
because Ulysses is several AU distant, these bursts can be precisely
localized to thin ($\sim$arc-minute) annuli through BATSE/Ulysses
arrival-time analysis.  Combination of the timing annuli with
independent BATSE location measurements results in precise
arc-segment-like uncertainty regions that are typically 25 times
smaller in area than the BATSE localizations alone.  In computing a
combined B/U burst localization, we model the BATSE location
uncertainty $B_{i}(\hat{r})$ with the ``core-plus-tail'' distribution
empirically determined by Briggs et al.\ (1998).  This is essentially
the weighted sum of two Gaussians ($\sigma_{1}=1.85\arcdeg$,
$\sigma_{2}=5.36\arcdeg$, and $f_{1}=0.78$), convolved with a
statistical uncertainty unique to each burst.  The timing location
uncertainty $U_{i}(\hat{r})$ is modeled with a spherical annulus of
Gaussian width unique for each burst.  The combined B/U localization
$P_{i}(\hat{r})$ is the product of $B_{i}(\hat{r}) \times
U_{i}(\hat{r})$, normalized over the unit sphere.

Of the 1637 events included in the fourth BATSE GRB catalog (4B[rev];
Paciesas et al.\ 1998), 415 were detected by Ulysses and have final
timing annuli given by Hurley et al.\ (1998) and Laros et al.\ (1998).
A map of the combined B/U localizations of these bursts is shown in
Figure 1, where, for display purposes, the localizations are
approximated with annular ring-segments (the actual shape is more
complicated).

%%\psdraft
\begin{figure}[]
  \centerline{ \psfig{figure=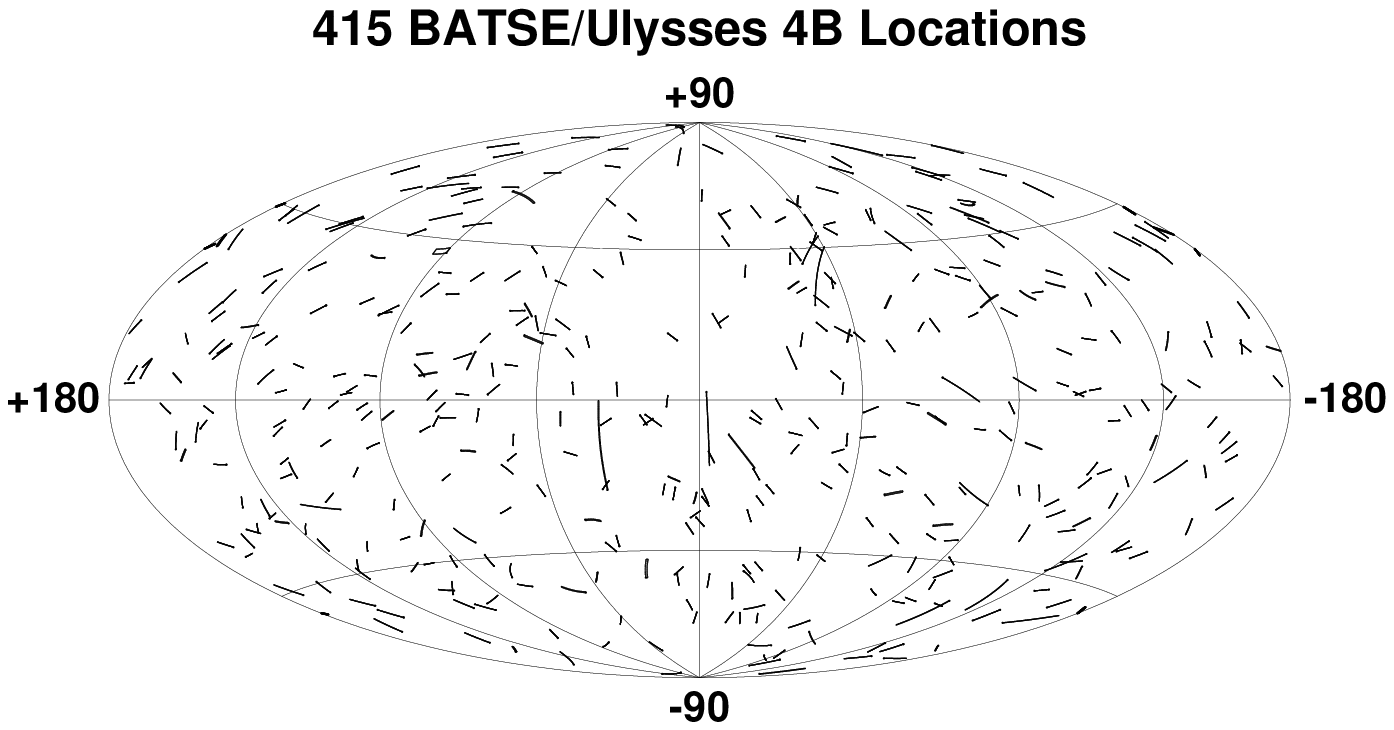,height=3.5cm} \hfill
               \psfig{figure=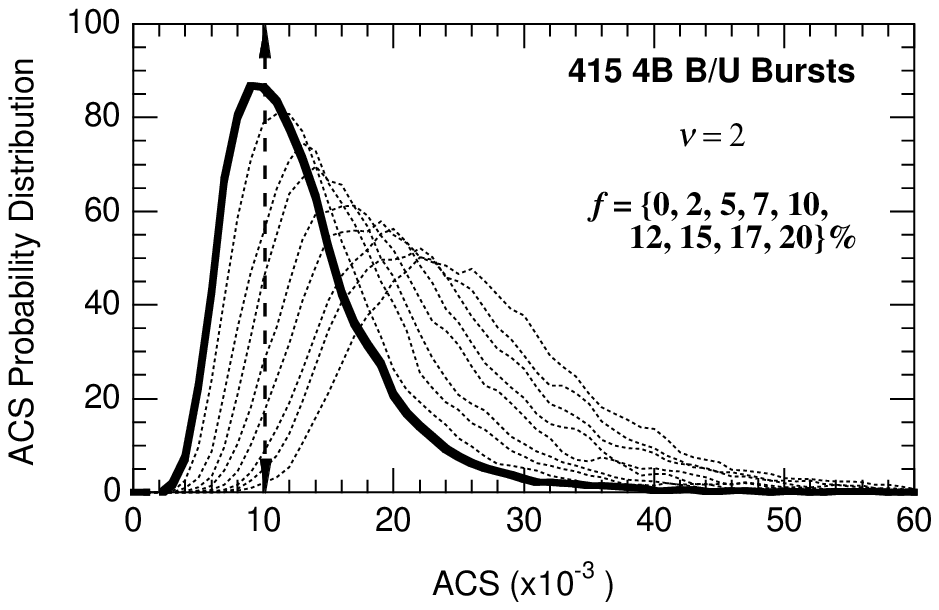,height=3.5cm} }
  \medskip
  \caption{FIG.\ 1. (left) Combined BATSE/Ulysses 4B burst locations 
               in Galactic coordinates.  (right) The measured ACS 
               statistic (vertical dashed line) compared to simulations
               of isotropic bursts with a fraction of clustered sources $f$
               ($\nu = 2$ detected bursts per cluster).}
\end{figure}

%-----------------------------------------------------------------------------%
\bsk
\ni 3. SPATIAL ANALYSIS TECHNIQUE
\ssk
\ni In earlier analyses of B/U burst locations, we used the traditional 
angular auto- and cross-correlation techniques to investigate
small-scale clustering and associations with known objects (Kippen et
al.\ 1998a, 1998b; Hurley et al.\ 1997).  These statistics suffer from
the fact that they depend on the user's choice of angular bin size.
For this paper, we have developed a more robust and more sensitive
tool, called the ``Arc Crossing Statistic'' (ACS), which is analogous
to the ``Total Power Statistic'' of Tegmark et al.\ (1996).

For a catalog of $N_{b}$ burst localizations $P_{i=1,N_{b}}$, the ACS
is defined as

\begin{equation}
  ACS \equiv {\left [  \sum_{i=1}^{N_{b}} \max{\{P_{i}(\hat{r})\}}^{2} 
      \right ]}^{-1} \sum_{i=1}^{N_{b}}\sum_{j=1}^{i-1} 
      \max{\{P_{i}(\hat{r}) \times P_{j}(\hat{r})\}}\ ,
\end{equation}

\ni where the term in square brackets is a convenient scale factor chosen
so that the result will be of order unity for an isotropic
distribution of sources with no clustering.  For investigating the
spatial association between GRBs and a catalog of $N_{x}$ objects with
known ``point'' locations $\hat{r_{j}}$, the statistic is changed to

\begin{equation}
  XACS \equiv {\left [ {N_{x} \over N_{b}} \sum_{i=1}^{N_{b}} 
      \max{\{P_{i}(\hat{r})\}} \right ]}^{-1}
      \sum_{i=1}^{N_{b}}\sum_{j=1}^{N_{x}} P_{i}(\hat{r_{j}})\ ,
\end{equation}

\ni where the leading term is chosen to yield XACS $\sim$ 1 for the case
of no association.  Note that ACS and XACS incorporate the actual
shapes of burst localizations in a continuous manner---eliminating the
need for any binning.  Also, because the $P_{i}$ burst localizations
are independently normalized, the relative contribution of each burst
to the ACS or XACS scales with the precision of their localizations.

To assess the significance of measured ACS/XACS values, we must
compare them to those expected from a random distribution.  This is
done via Monte Carlo simulations wherein random GRB location catalogs
are sampled from an isotropic angular distribution (corrected for
non-uniform observing exposure) and given the same uncertainties as
the real burst catalog.  The ACS/XACS from the random catalogs
provides a measure of the statistical distribution of expected values
and the significance $Q$ of the measured values $ACS_{obs}$ is given
by the fraction of simulated catalogs having $ACS
\geq ACS_{obs}$.

%-----------------------------------------------------------------------------%
\bsk
\ni 4. RESULTS
\ssk
\ni 4.1 Small-Scale Angular Clustering
\ssk
\ni Results from the ACS test applied to the catalog of 415 combined 
B/U 4B burst locations are shown in Figure 1.  It is evident that the
measured ACS value is consistent with the distribution expected for
isotropy.  Similar results are obtained for sub-catalogs of the most
precise locations, and for the sub-set of 3B data.  We therefore
investigate how much clustering is allowed by the data.  To do this we
model small-angle clustering with two standard parameters: the
fraction of bursts in ``point'' clusters ($f$) and the number of
detected bursts per cluster ($\nu$).  Figure 1 shows how the ACS
becomes increasingly inconsistent with the data as $f$ increases for
$\nu = 2$ (the most difficult case to detect).  Overall, we find that
the B/U 4B data require $f(\nu - 1) \lesssim 13\%$ at the 99\%
confidence level.
\ssk
\ni 4.2 Association with Extragalactic Objects
\ssk
\ni To examine possible associations between GRBs and known extragalactic
objects, we applied the XACS technique between the B/U 4B burst data
and various candidate source catalogs---the results are summarized in
Table 1.  These specific object catalogs were chosen because of past
reports of significant correlations with BATSE bursts.  As indicated
in the table, we find no significant correlation with any of the
catalogs (i.e., $Q$ is large).  By forcing a fraction of the bursts to
originate at known object locations, we can simulate the signal
expected from a true correlation.  The last column of Table 1 ($f_{\rm
max}^{99\%}$) is the largest fraction allowed by the data at the 99\%
confidence level.  Note that although we assume the published object
positions are exact points, the XACS technique is, by definition,
sensitive to cases where this may not be true.

%-----------------------------------------------------------------------------%
\bsk
\ni 5. CONCLUSIONS
\ssk
\ni Our results on small-scale clustering indicate no significant
deviations from isotropy and rule-out clustering and/or recurrence in
all but a small fraction of bursts.  Our limits may appear less
constraining than previous results (e.g., Tegmark et al.\ 1996), but
this is entirely because the earlier studies used BATSE location
uncertainties now known to be too small.  The combined B/U data
provide the most constraining limits on repetition and clustering.
These results have important implications if bursts are at large
redshift, where gravitational lensing is expected to result in
significant apparent small-angle clustering (Holtz, Miller \&
Quashnock 1999).

Our study also rules-out significant associations with any of the most
commonly known extragalactic objects.  The data indicate that only
very small numbers of bursts can possibly originate from the objects
tested.  These results are consistent with the currently favored
hypothesis that bursts originate from ``normal'' star forming galaxies
at redshift $z \sim 1$.

In conclusion, it should be duly noted that our results are all based
on high-fluence B/U bursts.  It is possible that clustering and/or
correlations could exist for a population of weaker events.  Sensitive
examination of this possibility could be accomplished by extending
our technique to the many weak bursts localized by BATSE, alone.

\begin{table}[]
  \centering
  \begin{tabular}[]{llccc}
    \multicolumn{5}{c} {TABLE 1. GRB Associations with Extragalactic Objects.}\\
    \hline
    \multicolumn{1}{c} {Objects} & {Number} & {Reference$^{\rm (a)}$} & {$Q$} & 
                       {$f_{\rm max}^{99\%}$ (\%)}\\
    \hline
    Abell Clusters (all)             &     5250 & Marani et al.\ (1997)   & 0.88 & 1.0 \\
    ' ' ($R \ge 1$, $D \le 4$)       & \phn 185 & Marani et al.\ (1997)   & 0.91 & 0.2 \\
    Radio Quiet QSO (all)            &     7146 & Schartel et al.\ (1997) & 0.10 & 3.9 \\
    ' ' ($z \le 1.0$, $M \le -24.2$) & \phn 967 & Schartel et al.\ (1997) & 0.13 & 1.2 \\
    AGN ($M_{\rm B} < -21$)          &     1390 & Burenin et al.\ (1998)  & 0.07 & 2.7 \\
    ' ' ($0.1 \le z \le 0.32$)       & \phn 543 & Burenin et al.\ (1998)  & 0.64 & 0.7 \\
    Recent SNe                       & \phn 599 & Kippen et al.\ (1998b)  & 0.94 & 0.2 \\
    \hline
    \multicolumn{5}{l} {\small $^{\rm (a)}$
       See references for details of catalog selection criteria.}
  \end{tabular}
\end{table}

%%%%%%%%%%%%%%%%%%%%%%%%%%%%%%%%%%%%%%%%%%%%%%%%%%%%%%%%%%%%%%%%%%%%%%%%%%%%%%%
\bsk
\baselineskip = 12pt
{\abstract \ni ACKNOWLEDGMENTS
\ni This research was supported, in part, through the Compton
Gamma Ray Observatory guest investigator program under grant
NAG5-6747.  
}

\bsk
\baselineskip = 12pt

%%%%%%%%%%%%%%%%%%%%%%%%%%%%%%%%%%%%%%%%%%%%%%%%%%%%%%%%%%%%%%%%%%%%%%%%%%%%%%%
% beginning of font "references"
{\references \ni REFERENCES
\ssk
\ref Briggs, M. S., et al.\ 1998, ApJS, in press
\ref Burenin, R. A., et al.\ 1998, Astron. Lett., in press
\ref Holtz, D. E., Miller, C. M., \& Quashnock, J. M. 1999, ApJ, in press
\ref Hurley, K., et al.\ 1998, ApJS, in press
\ref Hurley, K., et al.\ 1997, ApJ, 479, 113
\ref Kippen, R. M., Hurley, K., \& Pendleton, G. N. 1998a, in
   AIP Conf.\ Proc.\ 428, Gamma-Ray Bursts: 4th Huntsville Symposium,
   ed. C. A. Meegan, R. D. Preece, \& T. M. Koshut (New York: AIP),
   p.114
\ref Kippen, R. M., et al.\ 1998b, ApJ, in press
\ref Laros, J. G., et al.\ 1998, ApJS, in press
\ref Marani, G. F., et al.\ 1997, ApJ, 474, 576
\ref Paciesas, W. S., et al.\ 1998, ApJS, in press
\ref Schartel, N., Andernach, H. \& Greiner, J. 1997, A\&A, 323, 659
\ref Tegmark, M., et al.\ 1996, ApJ, 466, 757
}                      

%%%%%%%%%%%%%%%%%%%%%%%%%%%%%%%%%%%%%%%%%%%%%%%%%%%%%%%%%%%%%%%%%%%%%%%%%%%%%%%
\end{document}